# Neutron Reflectometry: a technique for revealing emergent phenomena at interfaces in heterostructures

Surendra Singh*

Solid State Physics Division, Bhabha Atomic Research Centre, Mumbai 400085 India

&

Homi Bhabha National Institute, Anushaktinagar, Mumbai 400094, India

*Email: surendra@barc.gov.in

Abstract:

Neutron reflectometry (NR) has emerged as a unique technique for the investigation of structure and magnetism of thin films of both biologically relevant and magnetic materials. The advantage of NR with respect to many other surface-sensitive techniques is its sub-nanometer resolution that enables structural characterizations at the molecular level. While in the case of bio-relevant samples, NR can be used to probe thin films at buried interfaces, non-destructively, even adopting a complex sample environment. Whereas the polarized version of NR is best suited for revealing the interface magnetism with a sub-nanometer depth resolution. In this article, I will briefly describe the basic principle of NR with some applications of NR to both bio-relevant samples and magnetic heterostructures.

# 1 Introduction

Surfaces and interfaces play a crucial role in defining the properties and functionality of both magnetic films [1, 2] and soft matter [3]. Materials at the interface often exhibit peculiar structure and behaviour, which are not observed in their bulk, as it encounters and interacts directly with different materials or phases through a very narrow region of the interface. While there are several practical phenomena relating to our daily life such as adhesion, lubrication, friction, coating, and painting, which are largely attributed to interfacial properties of materials in soft matter, emerging phenomena at interfaces of magnetic heterostructures and thin films attracted much attention for both industrial applications as well as academic researches. In addition, for solving complex biological problems and advancement in the development of highly functionalized biomimetic materials, understanding biological systems at a nanometer resolution is extremely important as many biological processes occur at interfaces in this length scale. Therefore, interface sensitivity has turned out to be the most useful characteristic in studies of magnetic thin films, multilayers, soft matters and biological systems. A wide range of direct and indirect techniques have been developed and used to probe surfaces and interfaces, including optical reflectivity/ellipsometry, optical imaging, fluorescence spectroscopy, surface probe techniques such as atomic force microscopy (AFM) and scanning tunnelling microscope (STM), the electron microscopy probes of the scanning electron microscope (SEM) and transmission electron microscopy (TEM), surface plasmon resonance (SPR), surface sensitive Fourier transform infrared (FTIR) spectroscopy, and X-ray reflectivity (XRR) and neutron reflectivity (NR). Each of these techniques suggests a particular viewpoint on the surface and interfacial properties. However, neutron scattering techniques such as NR offers unique property that makes them powerful techniques for the study of surface and interfaces.

In recent years, XRR and NR have emerged as particularly powerful techniques for probing surface and interfacial structures on a length scale of ~10 to 5000 Å [2]. The sensitivity of the NR technique to interfaces is because the projection of the neutron wavelength normal to the surface matches the thickness of thin films and the neutron wave function becomes strongly distorted near interfaces when the neutron encounters a potential step [2, 4-6]. NR probes the nuclear scattering length density (NSLD) profile perpendicular to the surfaces and interfaces (in contrast to the electron scattering length density, ESLD in XRR), the interface roughness, and the interface morphology [7]. In general, neutron scattering deals with the nuclear interaction of cold/thermal (energy $\leq$ 25 meV) neutrons and the strength of the interaction is characterized by the neutron scattering length, $b$, which varies randomly throughout the periodic table [8]. An important aspect of neutron scattering is that the value of b depends on isotopes and this is especially important for hydrogen (H) and deuterium (D or $^2$H), which have scattering lengths of different sign and magnitude ($b$ values for H and D is $-0.374 \times 10^{-12}$ cm and $0.6674 \times 10^{-12}$ cm, respectively) [8], thus it can be used to study biological systems. Apart from a few strongly adsorbing elements (e.g. Gd, Sm and Cd), the absorption cross-sections of cold/thermal neutrons are low, suggesting that neutrons are highly penetrating and nondestructive probes. Thus, the NR can probe buried interfaces such as solid/liquid interfaces nondestructively and also can make in situ measurements under various sample environments. In addition, neutrons are unique microscopic probes of magnetism, due to their inherent magnetic moment ($\mu_N$) of ~ -1.913 nuclear magneton (nuclear magneton = $-9.6623647 \times 10^{-27}$ J/T) that can examine the internal magnetic field (*B*) on the atomic level inside materials. Thus, using polarized NR (PNR) the magnetization depth profile is probed in addition to the nuclear depth profile of magnetic films and heterostructures.

Thus, surfaces and interfaces are present everywhere and the transport of atoms, molecules, or charge across interfaces is fundamentally important to the properties and function of materials. Also, the potential energy gradients at interfaces drive a host of thermodynamic processes. Emerging properties pursued through materials by design will be realized only in new materials having at least one nanoscale dimension e.g. composite materials and heterostructures. New functionality in these systems can arise from interface-dominated forces and interactions, rather than from the less-specific interactions that span the bulk. NR is being used to solve a variety of problems in the field of material science [4, 9, 10], polymers and soft matters [11-13], thin film magnetism [4, 14-19], superconductivity [20-24], chemistry and biology (e.g. surfactants, polymers, lipids, proteins, and mixtures adsorbed at liquid/fluid and solid/fluid interfaces) [25-28]. Here, I will briefly describe the principle of specular (angle of incidence = angle of reflection) neutron reflectivity using a few model systems containing single and multiple interfaces. Also, an experimental procedure with a neutron reflectometer instrument will be described briefly. A typical example of emerging interface magnetism in magnetic heterostructure and soft matter interfaces, where neutron reflectivity has played an important role, will also be described in this review.

## 2 Neutron Reflectometry

The reflection of light from surfaces, as a result of a change in refractive index, is a well-known phenomenon. Interference of light reflected from the front and back surfaces of the film and optical interference are still used to measure the thickness of surface coatings. During an experiment in the 1920s, Compton showed that X-ray reflection follows the same laws as a reflection of light but with different refractive indices and the refractive index for the X-ray depends on electron number density [29]. In 1944 Fermi and Zinn first demonstrated the mirror reflection of neutrons and neutron reflectivity (NR) measurement for finding out coherent nuclear scattering cross-section of various materials [30]. Most of the neutron scattering techniques, which are now commonplace for studying all branches of basic science and materials, were already well established in the early 1970s but not neutron reflectometry. However, NR emerged and rapidly developed as a dedicated technique in the later 1970s and mid-1980s, for the study of surfaces and interfaces. NR follows the same fundamental equations as optical reflectivity but with different refractive indices. A neutron refractive index of any material depends on the scattering length density of its constituent nuclei and the wavelength of the neutron. The neutron refractive index for most materials is only slightly less than that of air or vacuum and thus total external reflection is more commonly observed instead of the total internal reflection experienced with light. Over the year neutron reflection has emerged as a powerful technique for investigating the inhomogeneities across the interface like inhomogeneities in composition or magnetization [13, 31].

## 2.1 Unpolarized Neutron Reflectivity

Unpolarized NR is mostly used to investigate the depth profiling of structure (NSLD depth profiles) of nonmagnetic samples, biophysics and soft matters. This technique is being used for studies of surface chemistry (surfactants, polymers, lipids, proteins and mixtures adsorbed at liquid/fluid and solid/fluid interfaces) and other soft matter interfaces. The non-destructive feature of NR is very helpful to probe bio-relevant thin films at buried interfaces as well as using such samples with bulky sample environment equipment. In addition, the advancement in biomolecular deuteration enabled to probe certain structural features and to accurately resolve the location of chemically similar molecules within a thin film. The schematic geometry of neutron reflectivity is shown in Figure 1. The intensity of the neutron beam reflected at a glancing angle from a flat material surface depends upon the nature of the surface as well as the composition of the underlying matter. Figure 1 shows reflectivity geometry for the two types of possible reflections from a rough surface: (a) Specular reflection: when the angle of incidence ($\theta_i$) is equal to the angle of reflection ($\theta_f$) [$\theta_i = \theta_f = \theta$], and (b) Off-specular scattering: where the above equality is not maintained ($\theta_i \neq \theta_f$). Specifically, specular reflectivity can be analyzed to reconstruct laterally averaged compositional depth profile along the normal to the surface (z-direction in Fig. 1). On the other hand, at the angular position, in which $\theta_i \neq \theta_f$, around the specular reflection position weak off-specular scattering is observed originating from in-plane structures of the sample in the *x*-direction. Experimentally, the specular reflectivity is measured as a function of the momentum transfer wave vector, $Q$ ($= 4\pi \sin\theta/\lambda$ where λ is the wavelength of neutrons), as defined in Fig. 1.

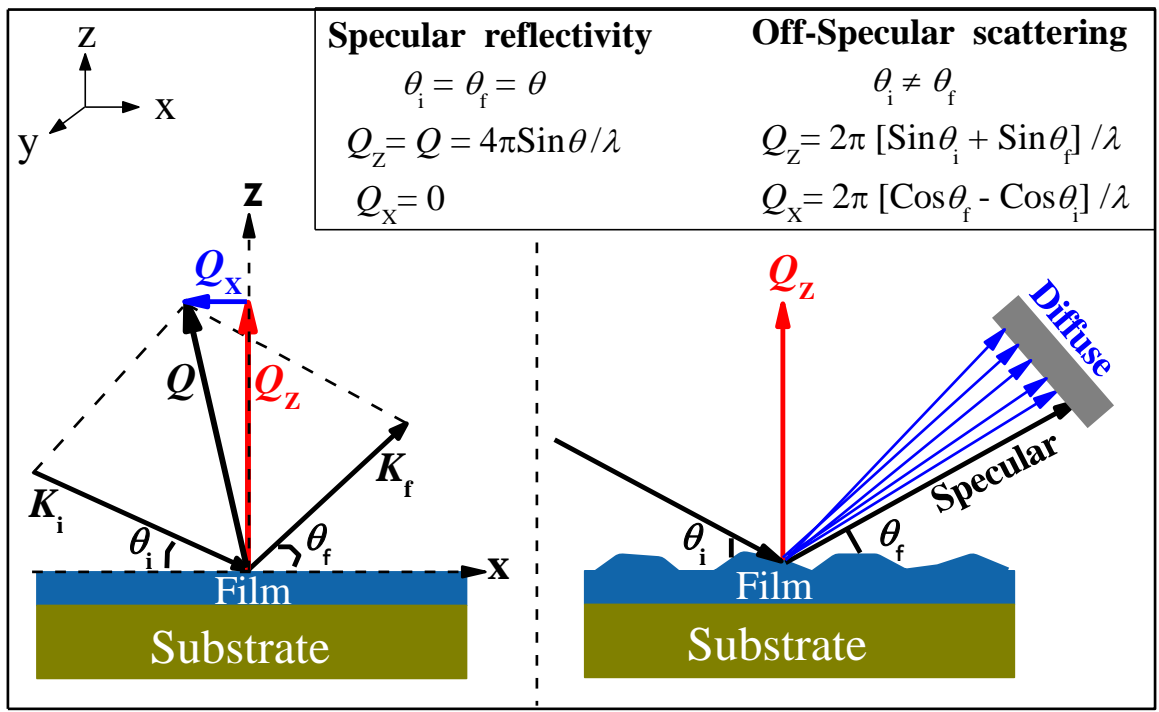

Figure 1: Geometry of specular neutron reflectivity and off-specular (diffuse) neutron scattering.

Neutrons, in the quantum mechanical approach, can be treated as a wave with a characteristic wavelength, $\lambda$, defined by the de Broglie relation:

$$\lambda = \frac{h}{m_n v} \qquad (1)$$

Where $h$, $m_n$ and $v$, are Planck's constant, the neutron mass and speed, respectively. The propagation of neutrons in any medium can be represented by the Schrödinger equation for the neutron wave function $\Psi(r)$ in the medium, which is analogous to the wave equation for light and leads to neutrons showing characteristic optical behaviour such as total reflection and refraction [32, 33]. The Schrödinger equation for the propagation of neutron beam in a medium can be written as:

$$-\frac{\hbar^2}{2m_n}\nabla^2\psi(\vec{r}) + V(\vec{r})\psi(\vec{r}) = E_k \psi(\vec{r}) \qquad (2)$$

Where $\hbar$ is Planck's constant divided by $2\pi$, $V$ is the potential seen by the neutron, $m_n$ is the mass of the neutron, $\vec{r}$ is the position vector of the neutron with wave function $\psi(\vec{r})$ and $E_k$ ($=\frac{\hbar^2 k^2}{2m_n}$, where, $k = 2\pi/\lambda$, is the wave vector of the neutron) its energy. In small angle limits, in which NR is measured and far from satisfying the Bragg condition of the crystalline structure, $V$ represents a constant potential (as depicted in Fig. 2), the net effect of the interactions between the neutron and the stutterers in the medium through which it moves and is simply related to the coherent scattering length by the relation [34]:

$$V = \frac{2\pi\hbar^2}{m_n}\rho \qquad (3)$$

Where $\rho$ is the NSLD of the medium and it is defined as $\rho = \sum_i N_i b_i$, with $N_i$ and $b_i$ are the number of nuclei per unit volume and scattering length of nucleus '$i$', respectively. The scattering length $b$, change not only from one atomic species to another but also for the different isotopes of the same species because the interaction of a neutron with a nucleus depends not only on the atomic number of the nucleus but also on the total spin state of the nucleus-neutron system. In general, the value of $b$ is a complex quantity and the imaginary part of $b$ ($=\sqrt{\sigma_a/4\pi}$, where $\sigma_a$ is absorption cross section for thermal neutrons) accounts for the absorption of the neutron in the medium. The absorption cross-section for neutrons is negligible for most of the elements, except for some elements e.g. Gd, Sm, B and Cd. Most of the elements (materials) in the periodic table have a positive $b$, which represents a positive potential for neutrons and thus has less kinetic energy. Neutrons seeing positive potential in the medium thus have a longer wavelength (opposite to light where the wavelength shortens) in the medium.

Consider a two-medium system, as depicted in Fig. 2 (a), with $k$ and $k_1$, which are the wave vectors of incident neutrons in these mediums. In the non-relativistic limit, the energy of the neutron in the medium is given by:

$$E_1 = \frac{\hbar^2 K_1^2}{2m}; E_2 = E_n - V = \frac{\hbar^2 K^2}{2m} - V \qquad (4)$$

The wave vector of the incident neutron $k$ changes to $k_1$ under influence of the nuclear potential. This allows us to correlate the refractive index of the medium, $n$, to the NSLD ($\rho$) of the medium as given below

$$n^2 = \frac{k_1^2}{k^2} = 1 - \frac{4\pi\rho}{K^2} = 1 - \frac{\lambda^2}{\pi}\rho \qquad (5)$$

For most of the materials, the refractive index for neutrons is marginally less than unity and $\frac{\lambda^2 \rho}{\pi}$ is typically in the range of 10$^{-6}$ and reflection of neutrons takes place at grazing incidence as stated earlier. Since the refractive index is less than unity, there will be a total external reflection for neutrons unlike optical rays (refractive index in the medium >1) where total internal reflection occurs. At the interface between two media, Snell's law applies i.e. for neutrons, $\cos\theta = n\cos\theta_1$. For the critical angle of incidence ($\theta_c$), we have the condition $\theta_1 = 0$. Then we have $\cos\theta_c = n$ and using Eqn. (5) we get the expression for $\theta_c$. i.e.,

$$\theta_c = \lambda\sqrt{\frac{\rho}{\pi}} \qquad (6)$$

For most of the materials, the critical angles are about a few arc minutes per Å wavelength. The corresponding momentum transfer vector for this angle of incidence is denoted $Q_c$ (= $4\pi\sin(\theta_c)/\lambda$).

If we consider a plane wave propagating in the *z-x* plane with the interface lying in the *x-y* plane and the stratification along the *z* direction (schematic is shown in Fig. 2 (a)), then the potential $V$ depends only on one spatial variable $z$. The solution of Eqn. (2) may be written as:

$$\psi(z,x) = e^{iKx}\phi(z) \qquad (7)$$

Using Eqns. (6) and (2) one gets a second-order differential equation for $\phi$:

$$\frac{d^2\phi}{dz^2} + q^2\phi = 0; q^2 = \frac{2m_n}{\hbar^2}[E_k - V] - K^2 \qquad (8)$$

Where $q$ and $K$ are $z$ and $x$ components of the wave vector $k$ respectively. Considering the unit incoming neutron ($i = 1$), the reflection amplitude ($r$) and the transmission amplitude ($t$) are defined in terms of the limiting forms of the solution of Eqn. (8):

$$e^{iq_1 z} + re^{-iq_1 z} \leftarrow \phi(z) \rightarrow te^{iq_2 z} \qquad (9)$$

Where $q_1$ and $q_2$ are defined in Eqn. (9) for two mediums (normal components of the wave vector in a vacuum and the medium, respectively), says, 1 and 2, which consist of an interface.

A typical two-medium system with an incident neutron beam ($i = 1$), reflected beam ($r$) and transmitted beam ($t$), is represented in Fig. 2(a).

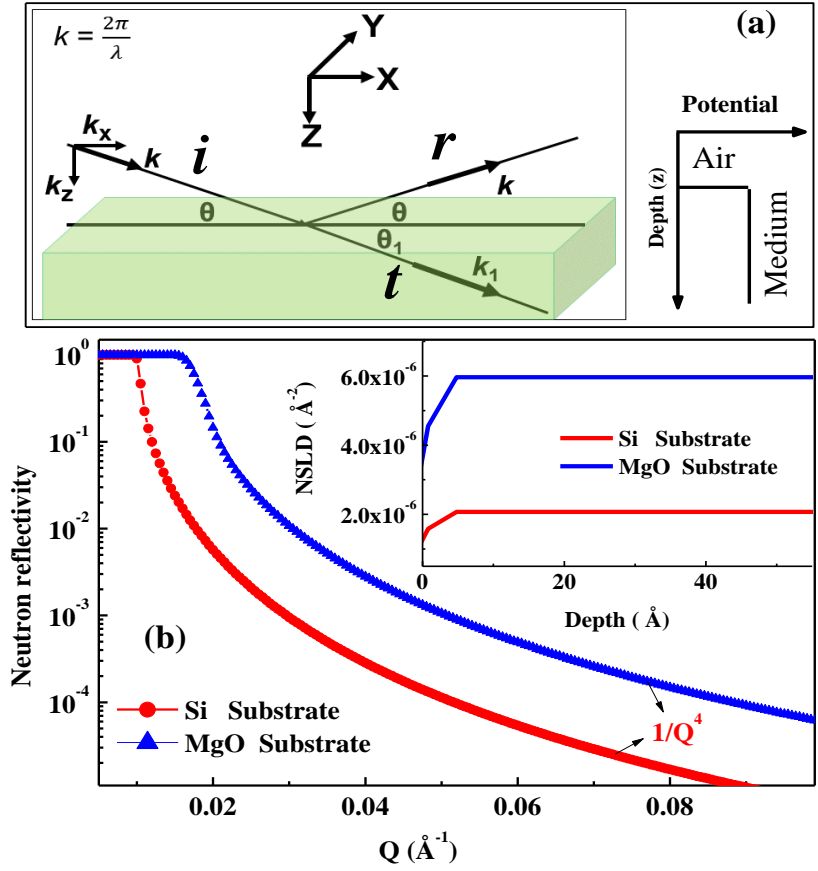

Figure 2: (a) Schematic of Neutron beam reflecting at sample surface with direction convention used. The $k$ and $k_1$ are the incident (medium 1) and refracted (medium 2) wave vectors, with angles θ (angle of incident = angle of reflection for specular NR) in the incidence plane. On the right, the depth profile of the potential of the medium is represented. (b) Comparison of calculated specular NR patterns from two substrates, Si and MgO. The inset shows the NSLD profiles for these substrates.

The intensity of the reflected specular signal from an ideally flat surface can be calculated by considering proper boundary conditions for neutron wave function $\Psi(z)$ and its derivative at the interface. The result is known as Fresnel relationships, which give the amplitude of specular reflection and the transmission coefficient of the beam. If one considers an ideally flat surface with a sharp boundary, then the reflection/refraction, at an interface, can be represented by Fig. 2(a). In this figure, the neutron beam impinges the surface from the vacuum at a glancing angle $\theta$, and the refracted beam propagates at an angle $\theta_1$. $k$ is the wave vector of the incident neutron. Considering the continuity of wave function $\phi$ (the solution of Eqn. (8)) and $\frac{d\phi}{dz}$ at the interface (i.e. Eqn. (9) ), we get :

$$1 + r = t; \quad q_1(1 - r) = q_2 t \tag{10}$$

Solving these two equations for *r* and *t*, we get the reflection amplitude and transmission amplitude at the interface [32]:

$$r = \frac{q_1 - q_2}{q_1 + q_2}; \quad t = \frac{2q_1}{q_1 + q_2} \tag{11}$$

Where the normal component of wavevector in two mediums are $q_1 = \frac{2\pi}{\lambda} \sin\theta$ and $q_2 = \sqrt{q_1^2 - 4\pi\rho}$. The Fresnel reflectivity for an ideally flat surface, for a glancing angle $\theta$, is, defined as

$$R_f = |r|^2 = \left|\frac{q_1 - q_2}{q_1 + q_2}\right|^2 = \left|\frac{\sin\theta - \sqrt{n^2 - \cos^2\theta}}{\sin\theta + \sqrt{n^2 - \cos^2\theta}}\right|^2 \tag{12}$$

From Eqn. (12), when $\cos\theta > n$ then *r* is a complex number and the Fresnel reflectivity is unity, i.e. for $\theta < \theta_c$ (critical angle of incidence) there will be a total external reflection of neutrons. Above the critical angle when $\theta \gg \theta_c$, the reflectivity drops off as $\theta^{-4}$ (or $Q^{-4}$, where $Q = 2q_1$) and Eqn. (12) reduces to $R_f \approx \frac{16\pi^2}{Q^4}\rho^2$. This rapid drop in intensity beyond the critical angle makes reflectivity experiment intensity limited at larger angles. Using the Eqn. (12) we have simulated the NR profile for Si and MgO substrates and it is shown in Fig. 2 (b). The inset of Fig 2(b) shows the NSLD depth profiles of these substrates. It is evident from Fig 2(b) that the reflectivity from the MgO substrate is higher as compared to the Si substrate, which is due to the larger NSLD value for MgO. Neutrons see a higher change in NSLD (contrast) at the air-substrate interface for the MgO substrate and thus are reflected neutrons strongly.

    Till now we have seen the NR from a single interface (air-substrate interface) consisting of two mediums. Calculation of NR for a single film grown on a substrate or a multilayer film grown on a substrate can be adopted following a similar approach and applying the boundary conditions at each interface of the heterostructures. The schematic of the heterostructure consisting of a single layer on a substrate and a periodic multilayer is shown in Fig. 3 (a) and (b), respectively, along with the potential depth profiles. When neutrons impinge on the film surface from the side of air from a single-layer heterostructure (Fig. 3(a)), interference occurs by a phase difference, corresponding to film thickness (*d*), between the neutrons reflected at the air-film interface (surface) and the film-substrate interface. General techniques such as the optical matrix method [35] or Parratt formalism [36] are used extensively to generate reflectivity profiles from heterostructures. Figure 3 (c) compares the specular NR profiles calculated for (i) a Si substrate, (ii) a thin film of deuterated polystyrene (dPS) with a thickness of 500 Å on a Si substrate, and (iii) a thin film of hydrogenated polystyrene (hPS) with thickness 500 Å on Si substrate. The profiles for the two polymer thin films exhibit regular oscillation due to interference of neutron beam reflected from air-film and film-substrate interfaces and these are called Kiessig fringes. The constant period of Kiessig oscillations

corresponding to the film thickness ($d$) and the film thickness can be estimated by measuring the frequency of the fringes ($\Delta Q$) using a relation $d = \frac{2\pi}{\Delta Q}$. The corresponding NSLD profiles for three different cases are shown in the inset of Fig. 3 (c).

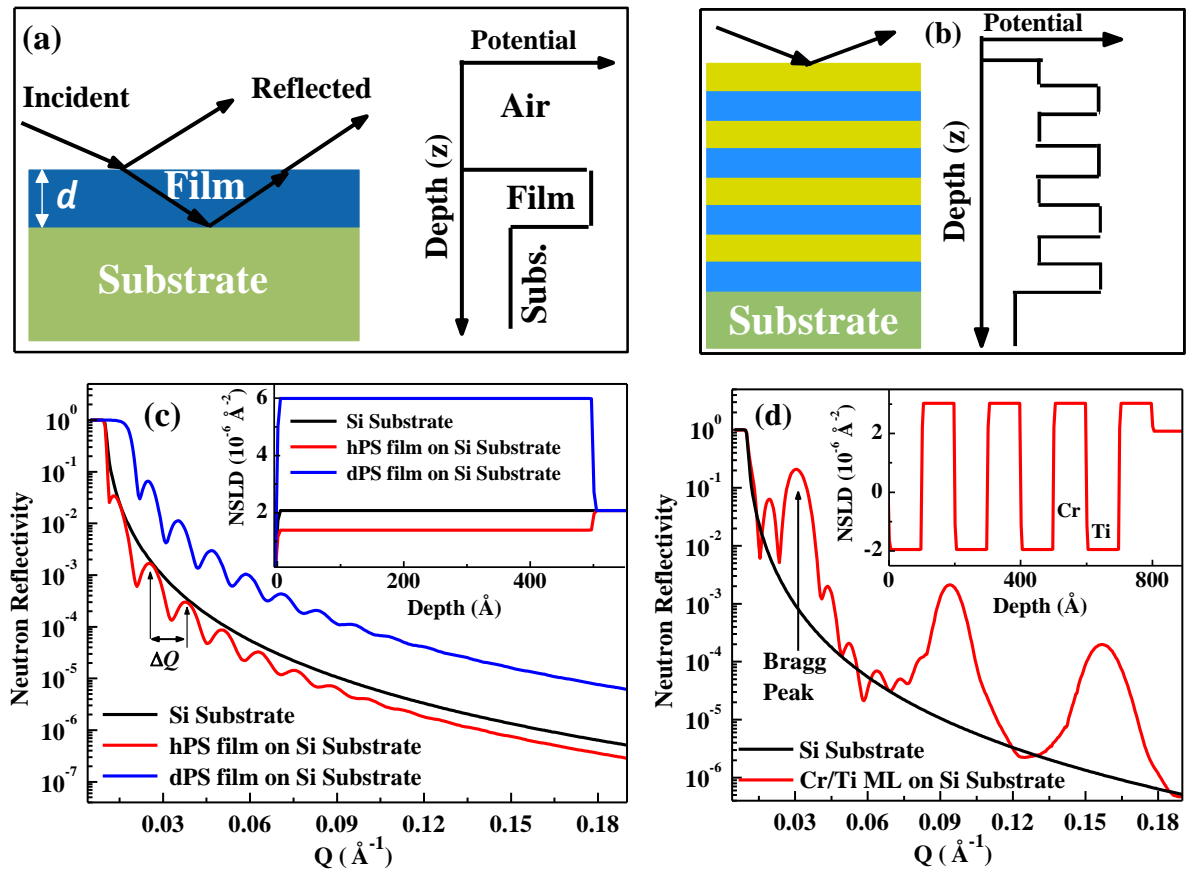

Figure 3: Schematic of neutron reflectivity and depth-dependent potential from (a) a single film and (b) a multilayer grown on a substrate. (c) comparison of calculated NR profiles from a Si substrate and a single film of the hPS (hydrogenated polystyrene) and the dPS (deuterated polystyrene) of similar thickness grown on a Si substrate. The inset shows the NSLD profiles. (d) comparison of calculated NR profiles for Si substrate and a multilayer of Cr/Ti (4 bilayers of Cr (~ 100 Å) and Ti (100 Å)) grown on Si substrate. The inset shows the NSLD depth profile of the multilayer.

It is noted that the neutron reflectivity for the dPS film is stronger than the hPS film, which is due to larger contrast (large change in NSLD at film-substrate and air-film interfaces) for the dPS film. A larger NR signal can help to collect the data for a larger Q range and this helps to estimate the parameters from the NR technique with a higher confidence limit. This example also suggests the importance of the deuteration of biological thin films for neutron reflectivity or in general neutron scattering techniques. Figure 3 (d) shows the NR pattern generated using Parratt's formalism [36] for a Cr/Ti multilayer of 4 bilayers with a bilayer period of 200 Å (thickness of 100 Å for each Co and Cu layers, grown on Si substrate). The inset of Fig. 3 (d) depicts the NSLD profile of the multilayer representing periodic NSLD variation in 4 bilayers. Thus using NR we can get a detailed layer structure of the designed multilayer.

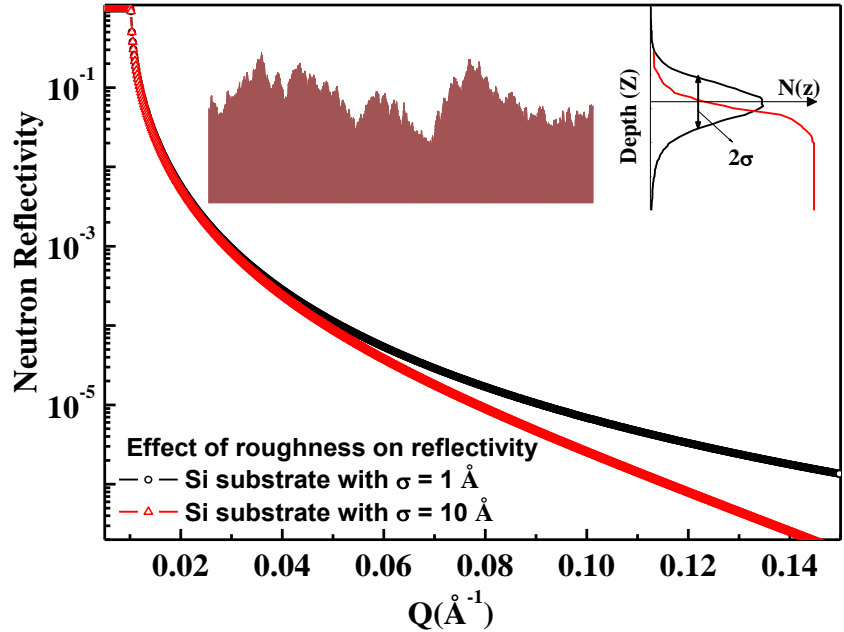

Figure 4: Effect of roughness on specular reflectivity from Si substrate with $\sigma = 0$ Å (solid line) and $\sigma = 10$ Å (dashed line). Inset (a) shows an image of a rough interface with a Gaussian profile of height. The standard deviation of the Gaussian function represents the root mean square roughness, $\sigma$.

Above, I have discussed the reflection from heterostructures with flat interfaces. However, there is a large interest in what happens between materials at an interface. An interface may be rough with peaks and troughs over a large range of length scales with a fractal-like structure. For magnetic heterostructure, there may also have magnetic roughness if the magnetization does not sharply change at an interface. A boundary may be smooth but with one material diffused into the other. The other kind of inhomogeneity, which arises at the interface, is the mixing of two materials due to inter-diffusion. It is very important to incorporate the roughness, i.e., the random character of the interface, into the reflectivity calculation. It turns out that in both the rough and inter-diffusion cases the specular reflectivity is reduced by a factor very much like the Debye-Waller factor reduces scattered intensity from a crystal [37]. The inset of Fig 4 shows a typical rough interface with the profile of the random height distribution, which is a Gaussian. The standard deviation of the Gaussian function represents the root mean square roughness, $\sigma$. The reflectance for a Gaussian rough surface is modified as [4, 7, 37]:

$$r(Q) = r_F e^{-2Q^2\sigma^2} \qquad (13)$$

Where, $r_F$ is Fresnel reflectance of the ideal surface given in Eqn. (12). Figure 4 shows the calculated NR profiles for Si substrates with a roughness of 1 Å (black open circles with line) and 10 Å (red open triangle with line). It is evident from Fig. 4 that the NR from a rough interface deviates from the ideal reflectivity for a smooth surface as a function of $Q$.

## 2.2 Polarized Neutron Reflectometry

Over the last 30 years, polarized neutron reflectometry (PNR) has contributed largely to investigating the interface magnetism in magnetic thin film heterostructures and magnetic multilayers. It is especially useful for functional materials and fundamentals of nanomagnetism, including exchange bias, multiferroic heterostructures, correlated electron systems, superconductors, magnetic nanoparticles and spin textures like skyrmions. Neutrons have a spin $\vec{\sigma}$, related to the magnetic moment $\vec{\mu}_n$ of the neutron by the vector operator: $\vec{\mu}_n = \mu_n \vec{\sigma}$, With $\mu_n$ = -1.913 $\beta_N$ and $\beta_N$ the nuclear magnetron ( = $e\hbar/2m_n c$). For a medium in which the scattering centres have magnetic moments, neutrons, because of their inherent magnetic moment, will experience an additional potential energy in a magnetic field $B$, other than the nuclear potential, given by:

$$V_{mag}(r) = -\vec{\mu}.\vec{B} \qquad (14)$$

Where $\vec{\mu}$ is the magnetic moment of the neutron and $\vec{B}$ is the magnetic field. It is clear that depending on the relative orientation of the neutron magnetic moment and the local field the magnetic potential (whether parallel or anti-parallel) can have a positive or negative value with respect to the nuclear potential. One may now write the total potential for the magnetic film as a summation of the nuclear and the magnetic parts as given below

$$V_{tot}(r) = \frac{2\pi\hbar^2}{m}(\rho_n \pm \rho_m) \qquad (15)$$

Where (+) and (-) signs correspond to the spin-up and spin-down neutrons with respect to sample magnetization. Using this potential one can solve the Schrödinger equation given in Eqn. (2) for calculation of the reflectivity from the magnetic medium in a similar way as given for unpolarized NR in the previous section. Now using Eqn. (5) and (6) the refractive index and critical angle for a neutron in a magnetic medium can be written as:

$$n = 1 - \frac{\lambda^2}{2\pi}(\rho_n \pm \rho_m); \qquad \theta_c = \lambda\sqrt{\frac{(\rho_n \pm \rho_m)}{\pi}} \qquad (16)$$

For calculating the PNR profile from a magnetic multilayer, there are formalism developed by C.F. Majkrazk [5] and G. P. Felcher [31], which describes the specific case of the neutron polarization axis being parallel to the sample surface. Parratt [36] formalism and the matrix formalism given by Blundell and Bland [35] are also used to generate the theoretical PNR profile and both give identical profiles.

Figure 5 shows a simulated polarized neutron reflectivity plot, of a Fe/Cr multilayer with 10 bilayers of Fe (thickness 60 Å) and Cr (50 Å) grown on Si substrate, as a function of wave vector transfer $Q$. For simulating the PNR profile, we have used the bulk magnetic moment (2.20 $\mu_B$) for Fe atom. The reflectivity pattern is generated using Parratt formalism [36]. The difference in the reflectivity pattern of the sample for the spin-up and spin-down neutrons is due to the difference in the step potential due to the magnetic part of the Fe layers for the spin-up and spins-down neutrons. The change in critical angle (Eqn. 16) for the two spins is also evident in Fig. 5.

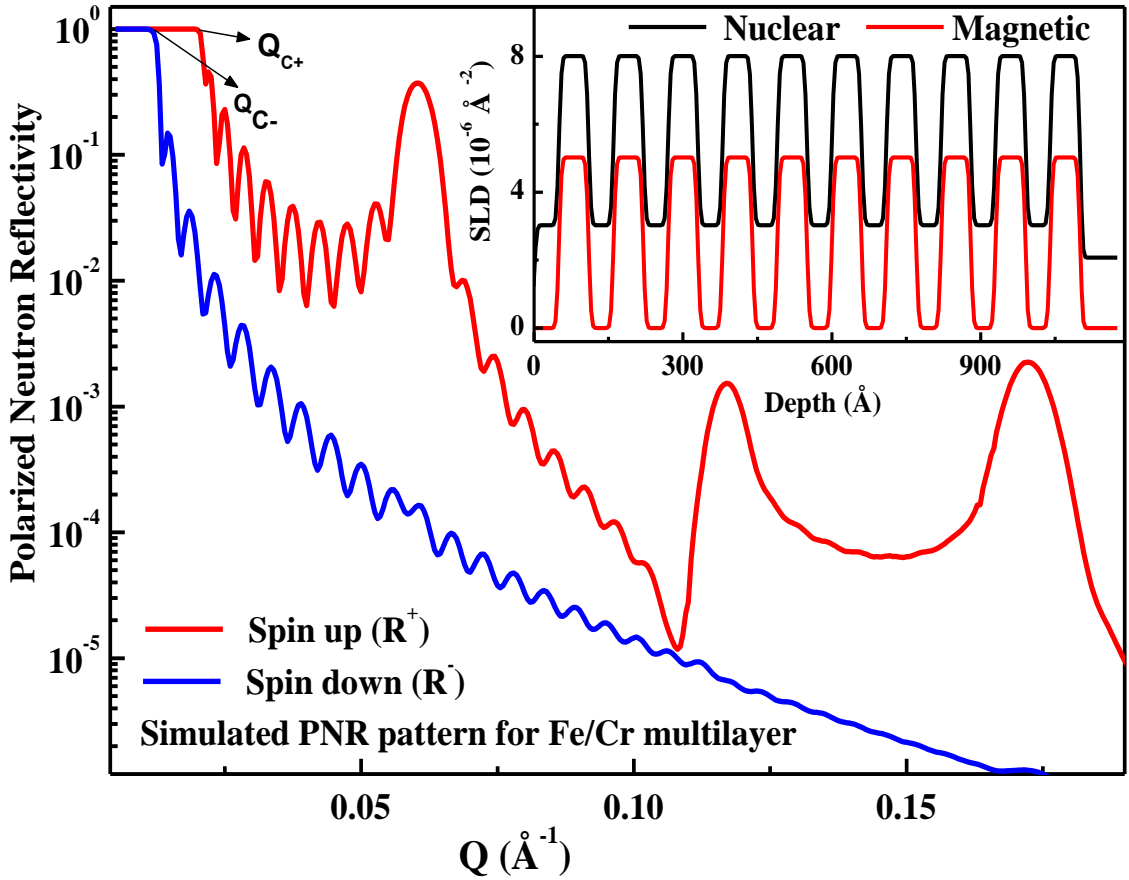

Fig. 5: Calculated polarized neutron reflectivity pattern for Fe/Cr multilayer.

In addition, if one needs to determine the in-plane magnetization vector (both magnitude and direction) and the nature of magnetic coupling in more complex magnetic multilayers the analysis of the polarization state of the reflected neutron beam is required. In this case, one obtains four reflectivity profiles with non-spin flip (NSF) ($R^{++}$, $R^{--}$) and spin-flip ($R^{+-}$, $R^{-+}$) modes. In the first approximation, the non-spin flip (NSF) components determine the component of the magnetic moment parallel to the quantization axis (or parallel to neutron spin) and the spin-flip (SF) components determine the perpendicular component. PNR in polarization analysis mode has provided valuable information on the nature of magnetic coupling in GMR multilayers and helped in understanding the nature of spin-dependent electron transport in these materials.

# 3  Neutron Reflectometer Instrumentation and Data Analysis

A number of neutron reflectometers are currently working at both types of neutron sources worldwide: a reactor and a spallation neutron source. In fact, every neutron facility in the world holds at least one reflectometer for carrying out structural studies on interfaces and thin films for a variety of materials. The conventional reflectometer, without a polarization option (for unpolarized NR measurements), has a relatively simple configuration, in which a pair of incident slits, a sample stage and a neutron detector are arranged in a sequence from neutron sources. The slit blades are made of a neutron-absorbing material such as cadmium, and the two incidents are separated from each other by approximately a few meters to produce a well-collimated beam falling on the sample table. The sample and the detector are mounted on goniometers or translation tables to make them arrange precisely at the proper angular position relative to the incident beam. While in the case of PNR measurements two additional components, polarizer and spin flipper are required to be added. Figure 6 shows the schematic of a PNR instrument. Depending on the type of neutron source (a monochromatic neutron beam usually from a reactor source or neutrons with a spread of wavelength usually from a spallation source) one can cover the reflectivity as a function of momentum transfer by changing the angle of incidence (reactor)/wavelength (spallation).

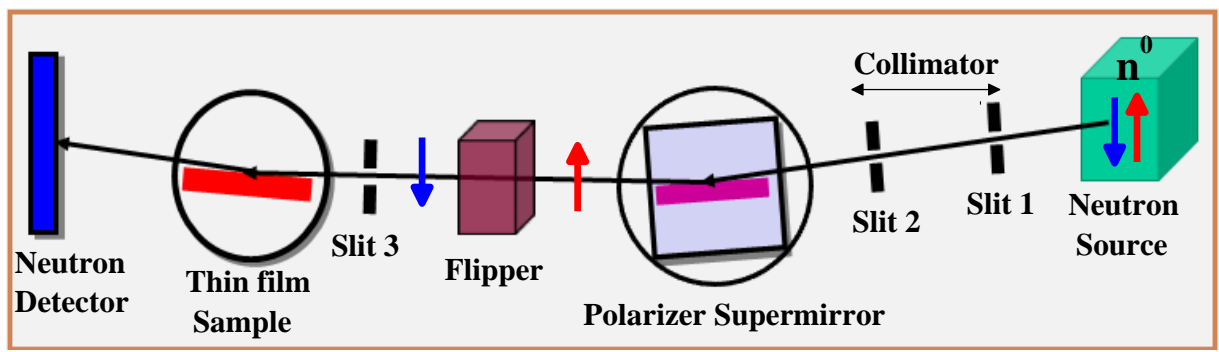

Figure 6: Schematic of a typical PNR instrument, which measures the spin-dependent reflectivity ($R^+$, spin-up and $R^-$, spin-down).

There are many neutron reflectometers available around the world in major neutron sources, which are being used for studying the interfaces. The full compilation of neutron reflectometers and the related data analysis program is available online [38]. A PNR instrument is available at the DHRUVA reactor, BARC, India [39, 40], which is designed for vertical sample geometry for PNR measurements without a polarization analysis option and is thus being used to investigate the magnetization depth profiles of thin films [4]. This instrument has been positioned on thermal neutron guide G2 in the Guide Tube Laboratory (GT Lab) of Dhruva Reactor, which uses a monochromatic neutron beam with a wavelength of $\lambda \sim 2.9$ Å and a linear position-sensitive detector (PSD), where only a sample is rotated to cover the desired Q range. Linear PSD also helps to collect both specular and off-specular reflectivity simultaneously at each angle of incidence [4, 39, 40].

In reflectometry, the reflected intensity is measured as a function of momentum transfer $Q$. The measured reflectivity is a convolution of reflectivity with the resolution of the instrument. To compare the measured data with the calculated data, first, we have to correct it for the instrumental resolution. This can be done either by, deconvoluting the measured data for experimental resolution or convoluting the theoretical profile with an experimental resolution before comparing it with experimental data. The instrumental resolution in small angle approximation is given by:

$$\frac{dQ}{Q} = \sqrt{\left(\frac{d\theta}{\theta}\right)^2 + \left(\frac{d\lambda}{\lambda}\right)^2} \qquad (17)$$

The typical wavelength resolution ($\frac{d\lambda}{\lambda}$) for a single crystal monochromator-based reflectometer is ~ 1 to 1.5%. The major contribution comes from the divergence of the neutron beam which can be varied by selecting different widths of slits and the typical $Q$ resolution ($\frac{dQ}{Q}$) of a reflectometer can be chosen around 1 to 3%. Usually, the reflectivity profile is measured for a constant value of $\frac{dQ}{Q}$ in the whole range of $Q$, which is done by choosing the appropriate combination of slits for various $Q$ ranges.

# 4 Interface Structures in Soft Matter Using Unpolarized Neutron Reflectivity

During the last decades, the use of unpolarized NR techniques in biological sciences has grown enormously [41]. This technique was introduced in the 1980s for the study of surfactant thin films and since then the improvements in neutron instrumentation, sample environment and modelling approaches have made this technique a critical tool for investigations of systems of increasing complexity. NR has emerged as a versatile tool for examining structural complexity at interfaces. This technique has been used to reveal the depth-dependent structure of biological samples including cell membranes, protein film, and a combination of protein and cell membranes, which is the most investigated class of the bio-relevant systems using NR. To understand the molecular processes involved in diseases and the development of new drugs, the interactions of proteins/peptides within and at the surface of cell membranes are essential [42]. As mentioned earlier, NR has been widely utilized for the investigation of depth profiling of structural properties on a variety of material interfaces and thin films for soft matters and biology: a polymer thin film, a Langmuir–Blodgett (LB) film, a Gibbs or a Langmuir monolayer of amphiphilic molecules such as surfactants, lipids, block copolymers, and proteins on the water surface, a polymer brush chemically or physically adsorbed on a solid substrate, and so on. Recently, NR in combination with X-ray reflectivity (XRR) was also being used to study the emerging field of liquid metals, organic liquids, liquid crystals, and ionic liquids for the understanding of the fundamental physics of liquid surfaces. Here, we will briefly describe the results on two systems, already published in litrature, highlighting the capability of NR.

## 4.1 Biological membranes

Bio-membranes are chemically diverse and complex molecular assemblies at the basis of life [43]. Cell membranes are active barriers for performing different and fundamental processes such as regulating cell growth, controlling the selective exchange of substances between inner and outer cellular environments and recognizing external stimuli [44]. Because of their nanometric thickness, they are perfect systems to be studied using NR [45]. Lipid bilayers are the basic building units of biological membranes and several biological processes involve revolutions between such bilayer structures, which are studied using unpolarized NR measurements. Phospholipid bilayers have been intensely used as model systems for studying the structure and interactions of biological membranes [46], which exhibit a main phase transition from a gel phase to a liquid crystalline phase [47]. In case of gel phase, lipid chains are rigid and well ordered, while in the liquid crystalline phase the chains are disordered and fluid like and such phase transition can be achieved by change the temperature of the system [47]. Such a structure phase transition in these biological model system can be studied using NR technique.

Figure 7 (a) shows a scematic of model system consisting of two bilayers (gel phase), one adsorbed on a flat surface of the Si substrate and a second floating at 20 to 30 Å above the first in bulk heavy water ($D_2O$). We have also shown the NSLD depth profile, which represents the variation of NSLD at different depth as per the structure of model lipid bilayer system in $D_2O$ medium, by blue line in Fig. 7 (a). In addition, we have also considered two more situations as shown in Fig. 7(b) and (c), where the bilayer lipid systems, shown in Fig. 7(a), show distortion from the gel phase. Figure 7 (b) shows the fluid phase where the lipid bilayer shows fluctuation in the position making the system with more roughness at different interfaces. The corresponding NSLD depth profile (blue solid line) is also depicted in Fig. 7 (b). Figure 7 (c) shows some intermediate phase where the separation between two bilayer also increases in addition to fluctuation in the top bilayer lipid. The corresponding NSLD profile is depicted in Fig. 7 (c). We have calculated the unpolarized NR profiles for these three situation and shown in Fig. 7(d), (e) and (f), respectively. IT is evident from Fig. 7 that the depth dependent fluctuation in this bio-interface system can be revealed using NR successfully. The similar depth dependent structure was also measured experimentally using NR for a hydrogenated di-stearyl phosphatidyl choline (DSPC) in $D_2O$ as a function of temperature [47].

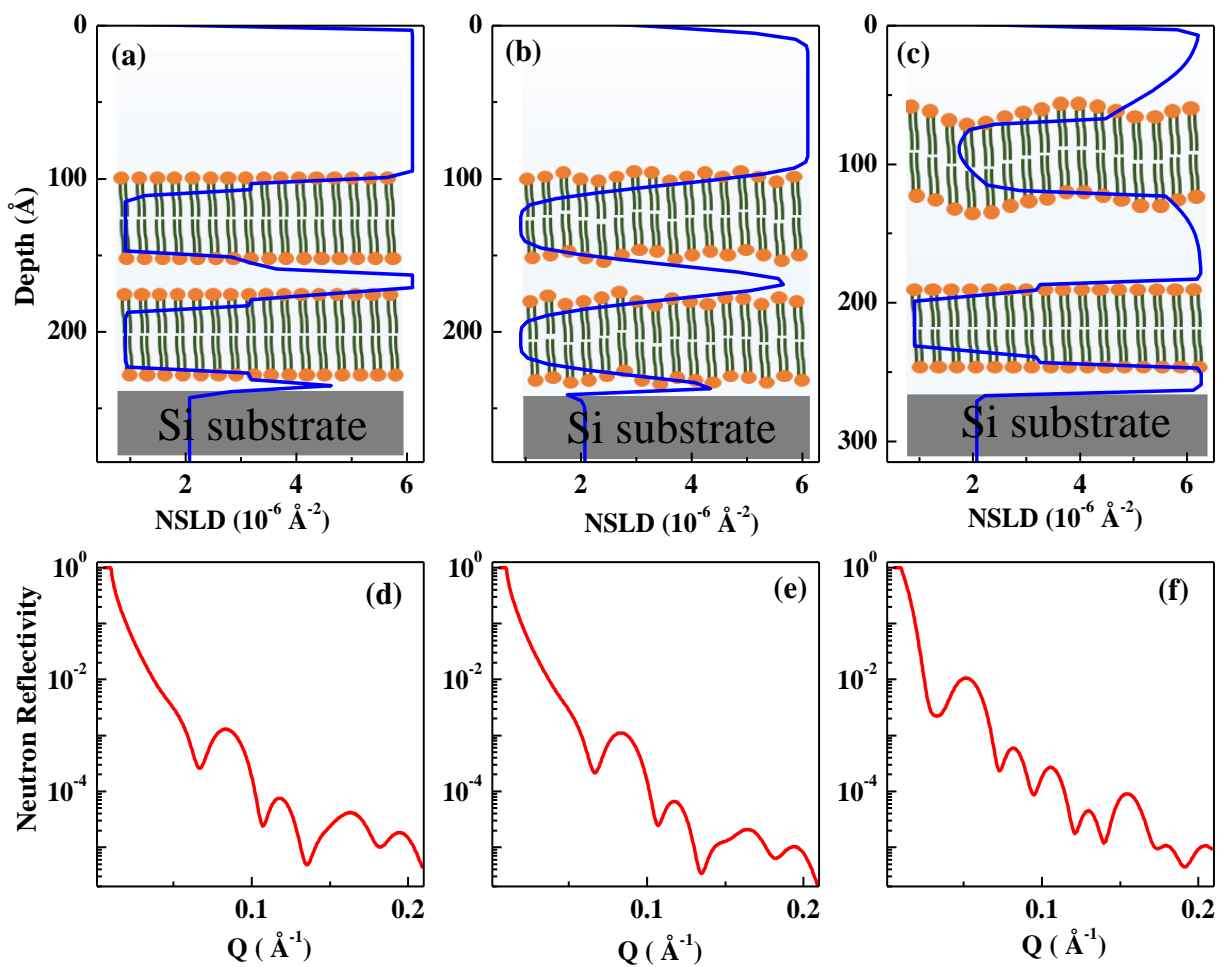

Figure 7: (a) Schematic model system of a biological membrane consisting of two lipid bilayers (gel phase), one adsorbed on a flat surface of the Si substrate and a second floating at 20 to 30 Å above the first in bulk heavy water ($D_2O$). The schematic of the two bilayer system with disorder is represented in (b) and (c). The Blue line shows the corresponding NSLD depth profiles. (d) to (f):The calculated NR patterns for three different models (shown in (a) to (c)).

### 4.2 Biomolecules at the interfaces and its interaction with membrane

Adsoption of biomolecules at interfaces, especially protein, provide an excellent opportunity and very active research area for which NR can contribute significantly. Though there are many techniques (e.g. ellipsometry, quartz crystal microbalance, atomic force microscopy etc.) which can be used to estimate the amount of absoerbed materials and related kineticks. NR is a characterization tool which provide detail organization of protein molecules at interfaces and it can uniquely study the exect location of protein molecules from the interfaces as well as formation of multilayers due to adsoption. Using NR, Mazzer *et al.* [48] studied the adsoption of immunoglobulin G (IgG) at various solid-liquid interfaces and NSLD depth profiles suggested the resistance of protein adsorption at the surfaces. Interaction of molecules with lipids permeate the daily life of human beings and NR allows to study these fundamental interaction not accessible by means of other techniques. The excellent use of NR in combination with other techniqes for studying the protein-membrane interactions is given by Martel et al. [49], where they have investigated the amyloidogenicity and cytotoxic mechanisms of an amyloid polypeptide in type 2 Diabetes Mellitus. Using NR they could quantify and localize the structural changes induced in a solid supported bilayer (membrane) by modeling the NR data with different depth profiling of NSLD values.

# 5 Emerging Phenomena at Interfaces of Magnetic Heterostructures Using Polarized Neutron Reflectivity

## 5.1 Non-magnetic Co film at Interfaces

3*d* transition metals e.g. Fe, Co and Ni show interesting magnetic properties both in their bulk and thin film form. Co in bulk shows a hexagonal closed pack (*hcp*) structure. However, the theory predicted that a high-density face-centred-cubic (*fcc*) phase of cobalt can be nonmagnetic [17, 50]. There was no experimental study in the literature for Co showing highly dense nonmagnetic properties. We investigated the depth-dependent structure and magnetic properties of a number of Co thin films (thickness of ~ 200-300 Å) deposited with different buffer layers on Si [111] single crystal substrate using PNR as a primary technique [17, 50]. PNR experiments in combination with other measurements confirmed the formation of a high-density Co layer at both interfaces (film-air or surface and substrate-film) of the film when Co was grown directly on Si layers. The PNR experiments were carried out at Dhruva. PNR results also suggested that the high-density interfacial Co layer is nonmagnetic. Figure 8 (a) and (b) show PNR data (symbols) as a function of the ratio of momentum transfer to the critical value of momentum transfer ($Q/Q_c$) along with fits (solid lines) from the Co film at room temperature (RT) grown on Si and $SiO_2$/Si substrates, respectively. Figure 8 (c) and (d) show the NSLD and MSLD depth profile of these heterostructures, which best fitted the PNR data at RT. The NSLD depth profile of the film obtained from the Co film grown on Si substrate (Fig. 8 (a)) indicated the presence of high-density (HD) Co layers at both interfaces [Fig. 8 (c)] i.e., film-air and film-substrate interfaces. Most interestingly, the HD Co layers at the interfaces were non-magnetic as can be seen in the MSLD (Fig. 8 (c)) depth profile obtained from PNR. Remarkably, when Co film was grown on $SiO_2$/Si substrate [Fig. 8 (b) and (d)] we did not obtain the HD Co layer formation at the interfaces.

It was also confirmed from high-resolution cross-sectional transmission electron microscopy that these high-density layers were of *fcc* structure unlike bulk Co, which has an *hcp* structure. Thus, it was conjectured that the migration of Co atoms along grain boundaries during deposition might have caused high pressure during the growth and given rise to such a phase at the interfaces, as the *fcc* phase can exist at high pressure. This unique result vindicated the strength of PNR in detecting such ultra-thin layers and their magnetism. These results demonstrate the fact that PNR plays an important role in obtaining such properties with excellent spatial resolution in sub-nanometer length scales.

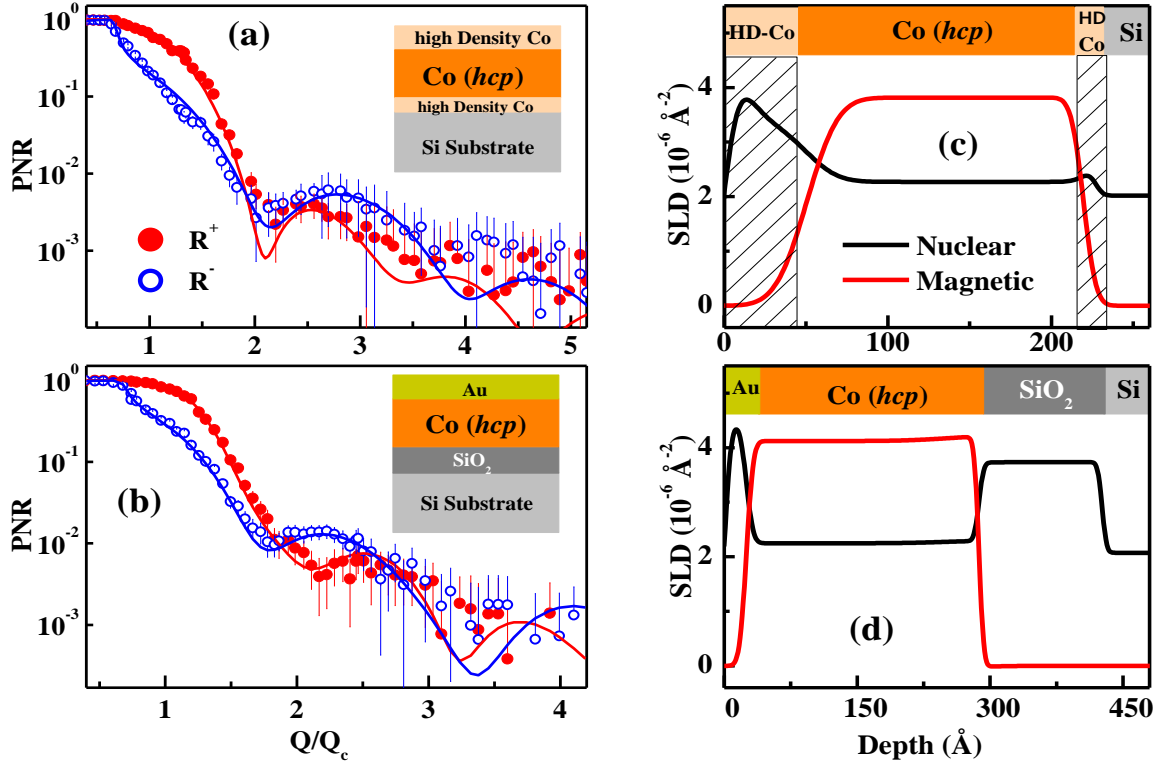

Figure 8: PNR data (symbol) and corresponding fits (solid lines) from Co films grown on (a) Si and (b) SiO$_2$/Si substrates. Insets (a) and (b) show a schematic of the corresponding film obtained from the PNR results. The NSLD and MSLD depth profiles obtained PNR results for Co films grown on (c) Si and (d) SiO$_2$/Si substrates. Co film grown on Si substrate suggested the formation of high density (HD) Co layer at both the interfaces which are non-magnetic.

## 5.2 Interface-Induced Magnetization in Co/Gd Multilayers

Rare earth (RE)–transition metal (TM) heterostructures, viz Gd-Fe, Gd-Co etc., are known to exhibit several magnetic structures at different temperatures and magnetic fields due to the strong antiferromagnetic (AF) interaction at the interfaces. Co and Gd show long-range magnetic order with a Curie temperature of ~ 1400 K and 293 K, respectively. Interface structure and morphology tend to play important roles in the magnetic properties of such systems. Using the polarized neutron reflectometer at Dhruva, we have investigated the structure and magnetic properties of Gd/Co multilayers at a nanometer length scale with an emphasis on the interface-induced phenomena. Interfacial dependence magnetic properties in this system were achieved either by depositing the multilayer using different growth parameters or annealing the multilayer samples [51, 52]. We have studied Gd/Co multilayer with a repeat of 10 bilayers of the Gd (thickness = 135 Å) and Co (thickness = 80 Å). Figure 9 (a) shows the schematic of a bilayer of Gd/Co before (as-deposited state, pristine) and after annealing the multilayer at a temperature of 400 °C for 0.5 hr. A comparison of macroscopic magnetization [$M(T)$], measured using SQUID magnetometer before and after annealing under field-cooled,

FC, (at an applied field of 500 Oe) and zero fields cooled (ZFC) are shown in Fig. 9 (b), suggesting temperature dependent modulation in the magnetization. Figure 9 (c) shows the PNR data (symbols) and corresponding fit (continuous lines) in the representation of spin asymmetry, defined as the ratio of the difference and sum of spin-dependent reflectivity $[(R^+ - R^-)/(R^+ + R^-)]$. Figure 9 (d) shows the magnetization depth profiles across the interfaces of a Gd/Co bilayer in the pristine and annealed states. PNR data suggested the formation of an alloy layer at the interface with reduced magnetization at both interfaces (Co/Gd and Gd/Co). This change in the interfacial structure and magnetic properties of the multilayer may be contributing to the modification in the exchange interaction resulting in a change in the macroscopic properties of the multilayer. Therefore, the PNR results as a function of interface structure and morphology in the Gd/Co multilayer systems [51, 52] offered valuable information to help us understand the mechanisms of interface-induced magnetization in RE/TM systems. These results also suggested that a more complicated model which takes into account interface roughness and morphology in RE/TM systems is required to study these systems micro-magnetically for possible future applications.

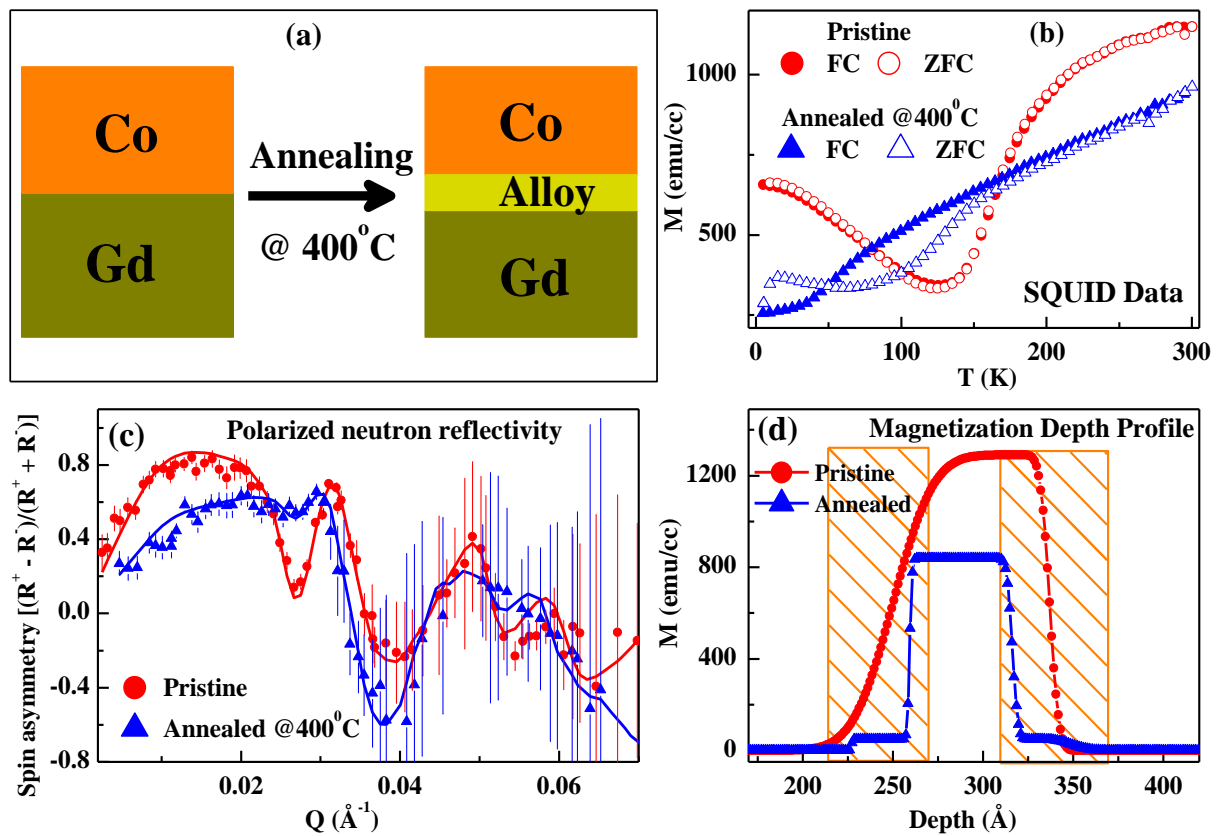

Figure 9: (a) Schematic of a bilayer of Gd/Co multilayer before and after annealing the multilayer. (b) $M(T)$ data in FC and ZFC mode from as-deposited (pristine) Co/Gd multilayer and multilayer annealed at 400 °C. (c) PNR data and corresponding fit from multilayer. (d) Magnetization depth profiles at two interfaces were obtained from PNR.

## 5.3 Proximity and Tunneling at interfaces in Superconductor and Ferromagnet Oxide Heterostructures

It is well-known that superconductors are perfect diamagnets that do not coexist with ferromagnets. This fact brings an interesting question to the fore: what happens if a superconductor (SC) and a ferromagnet (FM) are brought closure to each other? Moreover, recently this problem has attracted a large number of theoretical and experimental studies on a phenomenon known as the *"proximity effect"* [21-23, 53, 54], which showed suppression of magnetism on the FM side with the formation of a magnetic depleted (MD) layer below superconducting transition temperature ($T_{SC}$). Using the inherent capability of PNR, we have observed modulation in the interfacial magnetization of the FM layer below $T_{SC}$, even separating the FM layer from the SC layer by an insulator (I), which was described in terms of leakage of the spin-triplet Cooper pairs into the FM layer, leading to a modulation in the magnetism of the interfacial FM layer [21-24]. Here we describe the observation of suppression of the ferromagnetic order and the influence of ion irradiation and sequence of layer stacking on ferromagnetic order in $La_{2/3}Sr_{1/3}MnO_3$ (LSMO) layer separated from a $YBa_2Cu_3O_{7-\delta}$ (YBCO) layer by a thin oxide insulator $SrTiO_3$ (STO). The hybrid heterostructures were grown on a single crystal MgO substrate by pulsed laser deposition technique [21]. Combining structural and macroscopic magnetization measurements using XRD and SQUID, with PNR measurements we correlated the structural and magnetic properties across $T_{SC}$. Figure 10 shows the comparison of different measurements on 3 hybrid heterostructures shown in the left column.

The top panel of Fig. 10 shows the schematic and different measurements for the MgO/YBCO/STO/LSMO heterostructure. The middle panel of Fig. 10 shows different measurements from the same heterostructure after ion irradiation with 120 MeV $Ag^{15+}$ ions at a dose of $1 \times 10^{12}$ ions/cm$^2$. While the bottom panel of Fig 10 shows the schematic and measurements from the MgO/LSMO/STO/YBCO heterostructure, where we have reversed the sequence of the layer. The XRD data from all three systems have shown crystalline structure for the YBCO layer with highly textured in the [00l] direction [22-24]. A comparison of SQUID data for the heterostructures suggested that heterostructure MgO/YBCO/STO/LSMO showed a $T_{SC}$ of ~ 68 K (upper panel of Fig. 10), which reduces to 30 K (middle panel of Fig. 10) on ion irradiation and increases to 86 K (comparable to bulk value ~ 92 K for YBCO) on reversing the sequence of the layer growth. This change in $T_{SC}$ is directly correlated to the atomic structure of the YBCO layer. It is evident from Fig. 10 that upon ion irradiation the Bragg peak shifts to a lower angle (middle panel), whereas changing the growth sequence shifted the Bragg peak to a higher angle (lower panel), suggesting the expansion and compression of the lattice constant along *c* direction of YBCO layer on irradiation and sequence change, respectively. The PNR measurements were carried out on these heterostructures across $T_{SC}$, which revealed the detailed magnetization depth profiles at different temperatures. The magnetization depth profiles at 10 K from these heterostructures are shown at the extreme left of each panel Fig. 10, suggesting the formation of the MD layer at the STO/LSMO interface. The interfacial LSMO layer in the magnetization depth profiles is highlighted as a shaded region. A thicker MD layer of thickness ~ 35 Å was observed for the MgO/YBCO/STO/LSMO heterostructure (upper panel), which reduces to (a) zero Å upon ion irradiation of the heterostructure (middle panel) and (ii) to a thickness of ~ 20 Å for heterostructure with reversed sequence (lower panel) below $T_{SC}$. The formation of the MD layer below $T_{sc}$ in these heterostructures is attributed to the tunnelling of Cooper pairs across the insulator. This finding has demonstrated the need for PNR to understand the interplay of superconductivity and magnetism in such hybrid systems.

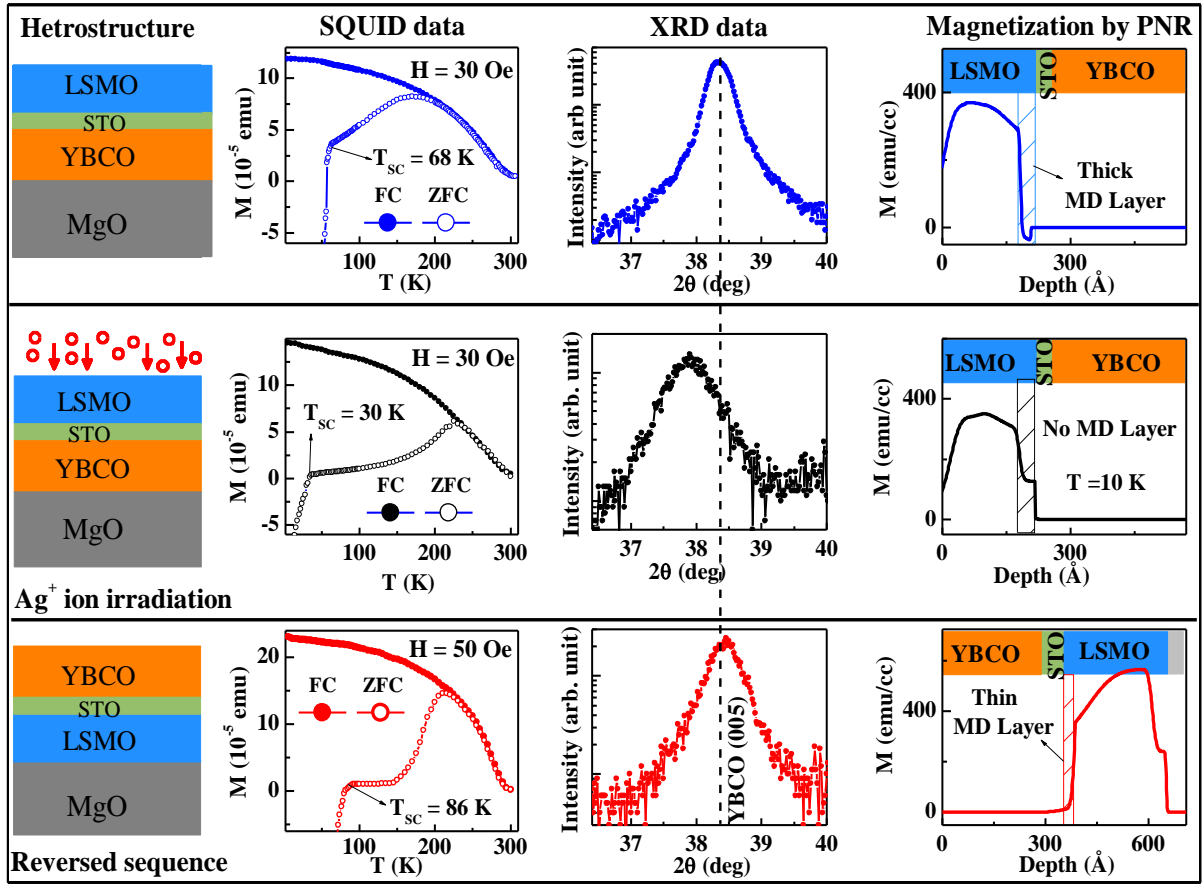

Figure 10: Comparison of XRD, SQUID data and magnetization depth profiles (PNR results) from trilayer heterostructures of YBCO-STO-LSMO grown on MgO substrates. (top panel) schematic of MgO/YBCO/STO/LSMO heterostructure with different measurements. (middle panel) ion irradiated MgO/YBCO/STO/LSMO heterostructure with different measurements. (lower panel) schematic of MgO/LSMO/STO/YBCO (reversed sequence) heterostructure with different measurements. The magnetization depth profiles extracted from PNR measurements for these heterostructures at 10 K are shown at the extreme right of each panel.

## 5.4 Induced Interface Magnetism at Ferromagnetic/Multiferroic Interface

The spin, charge and orbital reconstructions at interfaces in complex oxide heterostructures have driven several fascinating interfacial phenomena. Heterostructures involving oxide ferromagnets such as $La_{0.7}Sr_{0.3}MnO_3$ or $La_{0.67}Sr_{0.33}MnO_3$ (LSMO) and oxide antiferromagnet such as $BiFeO_3$ (BFO) is the one which has been explored extensively [16, 55-58]. BFO is a single-phase multiferroic material which exhibits magnetoelectric coupling between ferroelectric, FE, ($T_C$ = 1103 K) and antiferromagnetism, AFM, ($T_N$ = 643 K) order parameters [59]. Theoretical calculations have suggested a weak FM for BFO (~0.03 $\mu_B$) as a consequence of canting of the AFM structure due to the Dzyaloshinskii-Moriya interaction [60]. Experimentally, an enhanced magnetic moment (~ 0.75 $\mu_B$) in the BFO layer in BFO/LSMO heterostructure was observed right at the interface, which was attributed to Fe-Mn hybridization and orbital reconstruction [55]. PNR measurements from BFO/LSMO heterostructures have also suggested an enhanced magnetization (~ 0.75 $\mu_B$ to 1.5 $\mu_B$) for the Interfacial BFO layer of thickness ~6 to 8 unit cells and also suggested the presence of both types of magnetic coupling, FM (magnetic moment of interfacial BFO layer and LSMO are in same directions along the applied field) and AFM (moments are opposite, i.e. the magnetic moment of the interfacial BFO layer is opposite to applied field and LSMO layer) [16, 57, 58]. The different magnetic couplings in this system have also been confirmed by the first principal calculations and it was attributed to different surface terminations [16, 58].

Figure 11 (a) shows the PNR measurements at different temperatures for the LSMO/BFO heterostructure grown on the MgO substrate. Spin-dependent ($R^+$ and $R^-$) PNR data (symbol) at 300 K did not show any difference ($R^+$ and $R^-$ are the same) suggesting the system is nonmagnetic at this temperature and a detailed NSLD depth profile (Fig. 11 (b)) for the system was obtained from PNR data at 300 K. PNR measurements at low temperatures showed a splitting in $R^+$ and $R^-$ data, suggesting FM behaviour of the system and detail MSLD depth profiles obtained at different temperatures are shown in Fig. 11 (c). PNR data at 5 K suggested that the interfacial BFO layer (thickness ~ 35 Å) is ferromagnetic with the magnetization aligned opposite to the applied field and magnetization of the LSMO layer (AFM coupling). To justify the magnetization depth profile obtained from PNR at 5K we have also assumed different magnetization profiles and a comparison of fit to PNR data and different magnetization profiles are shown in Fig. 11 (d) and (e), respectively. It is evident from Fig. 11 (d) and (e) that the magnetization depth profile assuming AFM coupling between the LSMO layer and the interfacial BFO layer best fitted the PNR data at 5 K. We have also studied different heterostructures of LSMO/BFO using PNR and observed both FM and AFM coupling as depicted in Fig. 11 (f).

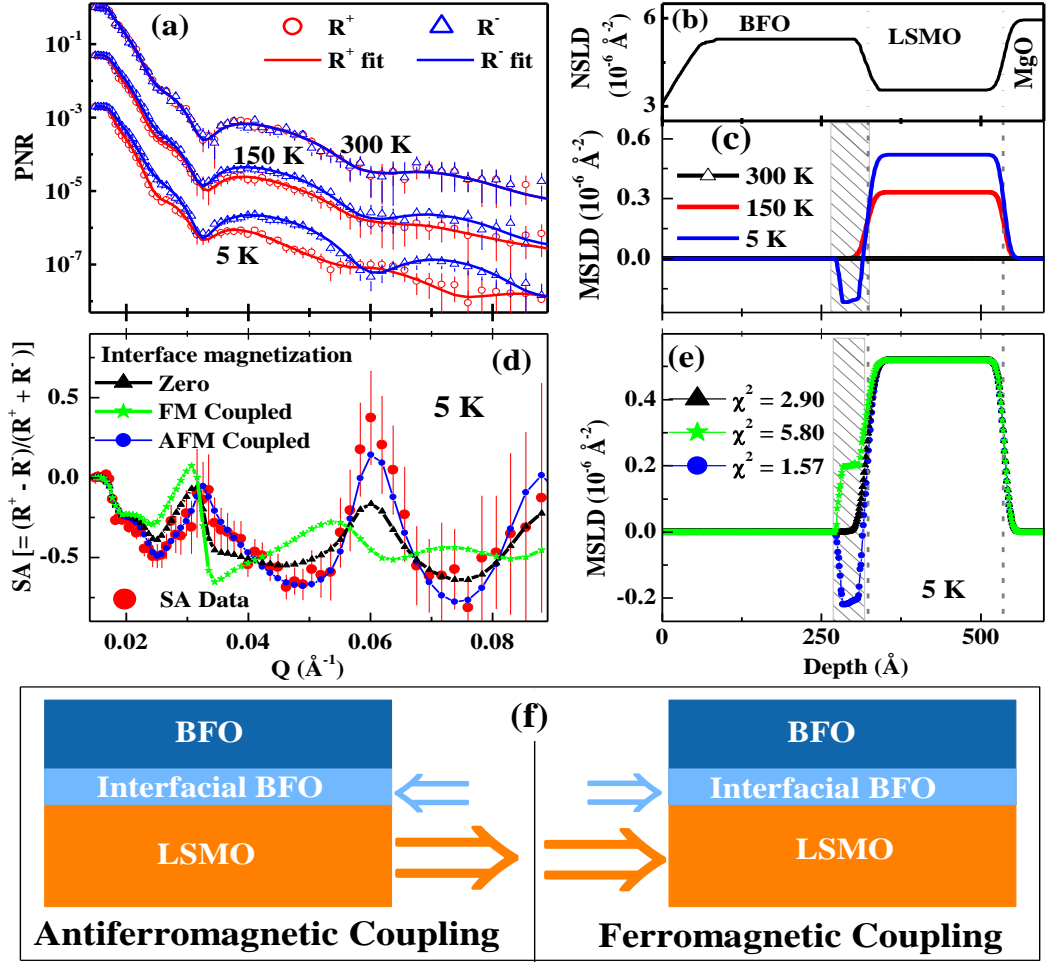

Figure 11: (a) PNR measurements at different temperatures from a MgO/LSMO/BFO heterostructure. The NSLD (b) and MSLD (c) depth profiles of the system were obtained from PNR at different temperatures. (d) Comparison of fits to spin asymmetry (SA) data (PNR data) considering different magnetization depth profiles as shown in (e). (f) different magnetic coupling between the LSMO layer and the interfacial BFO layer.

## 5.5 Other Examples of Emerging Interfacial Magnetism using PNR

The existence of different magnetic orders at interfaces, which are different from their constituent bulk phase, have been reported, e.g., ferromagnetism at interfaces between two different AFM oxides [61-63] and between a paramagnetic metal and AFM insulator oxides [64, 65], a ferromagnetic interfacial layer of paramagnetic oxide material between paramagnetic and ferromagnetic oxides [19] and ferromagnetism for a topological insulator (TI) interfacial layer between TI and ferromagnetic oxide [66]. These are the systems, where PNR was used as a unique probe to explore and understand such emerging phenomena at the interfaces. Recently, NR technique, due to its intrinsic strength in deciphering different isotope and interface magnetism, has played an important role in revealing the interfacial structure and magnetic phases as well as its coupling in different heterostructures.

# 6 Summary

In summary, I briefly discussed the specular NR technique with both unpolarized and polarized modes and its successful use for the investigation of emerging phenomena at interfaces of both soft (biophysical and soft matter) and hard (magnetism) condensed matter physics. It is fair to stress that neutron reflectivity, due to its inherent sensitivity to reveal small nuclear and magnetic structure with sub-nanometer depth resolution, has been and will continue to play a unique and major role in this exciting research area of thin films and heterostructures.

# References

[1] A.O. Adeyeye, G. Shimon, Chapter 1 - Growth and Characterization of Magnetic Thin Film and Nanostructures, in: R.E. Camley, Z. Celinski, R.L. Stamps (Eds.) Handbook of Surface Science, North-Holland 2015, pp. 1-41.

[2] B.P. Toperverg, H. Zabel, Neutron scattering in nanomagnetism, in: F. Fernandez-Alonso, D.L. Price (Eds.) Neutron scattering—magnetic and quantum phenomena, Experimental Methods in the Physical Sciences, Academic Press 2015, pp. 339-434.

[3] D. Zang, Y. Tarasevich, L. Zhang, S. Tarafdar, C. Ding, Self-Assembled and Artificial Surfaces/Interfaces: From Soft Matter to Metamaterials, Advances in Condensed Matter Physics, 2018 (2018) 9701423.

[4] S. Singh, M. Swain, S. Basu, Kinetics of interface alloy phase formation at nanometer length scale in ultra-thin films: X-ray and polarized neutron reflectometry, Progress in Materials Science, 96 (2018) 1-50.

[5] C.F. Majkrzak, Neutron scattering studies of magnetic thin films and multilayers, Physica B: Condensed Matter, 221 (1996) 342-356.

[6] M.R. Fitzsimmons, C. Majkrzak, Modern Techniques for Characterizing Magnetic Materials,, Springer, New York2005.

[7] S.K. Sinha, E.B. Sirota, S. Garoff, H.B. Stanley, X-ray and neutron scattering from rough surfaces, Physical Review B, 38 (1988) 2297-2311.

[8] A.J. Dianoux, G.H. Lander, Neutron Data Booklet, Old City 2002.

[9] S. Singh, S. Basu, P. Bhatt, A.K. Poswal, Kinetics of alloy formation at the interfaces in a Ni-Ti multilayer: X-ray and neutron reflectometry study, Physical Review B, 79 (2009) 195435.

[10] S. Singh, S. Basu, M. Gupta, C.F. Majkrzak, P.A. Kienzle, Growth kinetics of intermetallic alloy phase at the interfaces of a Ni/Al multilayer using polarized neutron and x-ray reflectometry, Physical Review B, 81 (2010) 235413.

[11] C. Carelli, R.N. Young, R.A.L. Jones, M. Sferrazza, The contrast match neutron reflectivity technique for the study of broad polymer/polymer interfaces, Nuclear

Instruments and Methods in Physics Research Section B: Beam Interactions with Materials and Atoms, 248 (2006) 170-174.

[12] T. Widmann, L.P. Kreuzer, G. Mangiapia, M. Haese, H. Frielinghaus, P. Müller-Buschbaum, 3D printed spherical environmental chamber for neutron reflectometry and grazing-incidence small-angle neutron scattering experiments, Review of Scientific Instruments, 91 (2020) 113903.

[13] J. Penfold, Neutron reflectivity and soft condensed matter, Current Opinion in Colloid & Interface Science, 7 (2002) 139-147.

[14] S. Singh, H. Bhatt, Y. Kumar, C.L. Prajapat, A. Mishra, S. Bedanta, S. Basu, Interface-driven static and dynamic magnetic properties of ultrathin Fe/Ge multilayers, Applied Surface Science, 570 (2021) 151193.

[15] S. Singh, M.R. Fitzsimmons, T. Lookman, J.D. Thompson, H. Jeen, A. Biswas, M.A. Roldan, M. Varela, Magnetic Nonuniformity and Thermal Hysteresis of Magnetism in a Manganite Thin Film, Physical Review Letters, 108 (2012) 077207.

[16] S. Singh, J.T. Haraldsen, J. Xiong, E.M. Choi, P. Lu, D. Yi, X.D. Wen, J. Liu, H. Wang, Z. Bi, P. Yu, M.R. Fitzsimmons, J.L. MacManus-Driscoll, R. Ramesh, A.V. Balatsky, J.-X. Zhu, Q.X. Jia, Induced Magnetization in $La_{0.7}Sr_{0.3}MnO_3/BiFeO_3$ Superlattices, Physical Review Letters, 113 (2014) 047204.

[17] N. Banu, S. Singh, B. Satpati, A. Roy, S. Basu, P. Chakraborty, H.C.P. Movva, V. Lauter, B.N. Dev, Evidence of Formation of Superdense Nonmagnetic Cobalt, Scientific Reports, 7 (2017) 41856.

[18] M.A. Basha, C.L. Prajapat, H. Bhatt, Y. Kumar, M. Gupta, C.J. Kinane, J.F.K. Cooper, A. Caruana, M.R. Gonal, S. Langridge, S. Basu, S. Singh, Helical magnetic structure and exchange bias across the compensation temperature of Gd/Co multilayers, Journal of Applied Physics, 128 (2020) 103901.

[19] H. Bhatt, Y. Kumar, C.L. Prajapat, C.J. Kinane, A. Caruana, S. Langridge, S. Basu, S. Singh, Emergent Interfacial Ferromagnetism and Exchange Bias Effect in Paramagnetic/Ferromagnetic Oxide Heterostructures, Advanced Materials Interfaces, 7 (2020) 2001172.

[20] J.M. Reynolds, V. Nunez, A.T. Boothroyd, R. Somekh, D.G. Bucknall, S. Langridge, Polarised neutron reflectometry studies of flux penetration in superconducting Nb, Physica B: Condensed Matter, 241-243 (1997) 1104-1106.

[21] C.L. Prajapat, S. Singh, A. Paul, D. Bhattacharya, M.R. Singh, S. Mattauch, G. Ravikumar, S. Basu, Superconductivity-induced magnetization depletion in a ferromagnet through an insulator in a ferromagnet–insulator–superconductor hybrid oxide heterostructure, Nanoscale, 8 (2016) 10188-10197.

[22] S. Singh, H. Bhatt, Y. Kumar, C.L. Prajapat, B. Satpati, C.J. Kinane, S. Langridge, G. Ravikumar, S. Basu, Superconductivity-driven negative interfacial magnetization in $YBa_2Cu_3O_{7-\delta}/SrTiO_3/La_{0.67}Sr_{0.33}MnO_3$ heterostructures, Applied Physics Letters, 116 (2020) 022406.

[23] Y. Kumar, H. Bhatt, C.L. Prajapat, A.P. Singh, F. Singh, C.J. Kinane, S. Langridge, S. Basu, S. Singh, Suppression of the superconducting proximity effect in ferromagnetic-superconducting oxide heterostructures with ion-irradiation, Journal of Applied Physics, 129 (2021) 163902.

[24] H. Bhatt, Y. Kumar, C.L. Prajapat, C.J. Kinane, A. Caruana, S. Langridge, S. Basu, S. Singh, Correlation of Magnetic and Superconducting Properties with the Strength of the Magnetic Proximity Effect in La$_{0.67}$Sr$_{0.33}$MnO$_3$/SrTiO$_3$/YBa$_2$Cu$_3$O$_{7-\delta}$ Heterostructures, ACS Applied Materials & Interfaces, 14 (2022) 8565-8574.

[25] S. Krueger, Neutron reflection from interfaces with biological and biomimetic materials, Current Opinion in Colloid & Interface Science, 6 (2001) 111-117.

[26] A. Tummino, E. Scoppola, G. Fragneto, P. Gutfreund, A. Maestro, R.A.W. Dryfe, Neutron reflectometry study of the interface between two immiscible electrolyte solutions: Effects of electrolyte concentration, applied electric field, and lipid adsorption, Electrochimica Acta, 384 (2021) 138336.

[27] W. Lu, J. Zhang, J. Xu, X. Wu, L. Chen, In Situ Visualized Cathode Electrolyte Interphase on LiCoO2 in High Voltage Cycling, ACS Applied Materials & Interfaces, 9 (2017) 19313-19318.

[28] J. Penfold, R.K. Thomas, The application of the specular reflection of neutrons to the study of surfaces and interfaces, Journal of Physics: Condensed Matter, 2 (1990) 1369-1412.

[29] T.H. Laby, R.T.W. Bingham, The reflection and diffraction of X-rays, Proceedings of the Royal Society of London. Series A, Containing Papers of a Mathematical and Physical Character, 133 (1931) 274-291.

[30] E. Fermi, W. Zinn, Reflection of neutrons on mirrors., Phys. Rev., 70 (1946) 103A.

[31] G.P. Felcher, R.O. Hilleke, R.K. Crawford, J. Haumann, R. Kleb, G. Ostrowski, Polarized neutron reflectometer: A new instrument to measure magnetic depth profiles, Review of Scientific Instruments, 58 (1987) 609-619.

[32] J. Lekner, Theory of reflection of electromagnetic and particle waves. , Martinus Nijhoff Publishers,1987.

[33] V.F. Sears, Fundamental aspects of neutron optics, Physics Reports, 82 (1982) 1-29.

[34] H. Zabel, X-ray and neutron scattering at thin films, in: U. Rössler (Ed.) Festkörperprobleme 30: Plenary Lectures of the Divisions Semiconductor Physics Thin Films Dynamics and Statistical Physics Magnetism Metal Physics Surface Physics Low Temperature Physics Molecular Physics of the German Physical Society (DPG), Regensburg, March 26 to 30, 1990, Springer Berlin Heidelberg, Berlin, Heidelberg, 1990, pp. 197-217.

[35] S.J. Blundell, J.A.C. Bland, Polarized neutron reflection as a probe of magnetic films and multilayers, Physical Review B, 46 (1992) 3391-3400.

[36] L.G. Parratt, Surface Studies of Solids by Total Reflection of X-Rays, Physical Review, 95 (1954) 359-369.

[37] L. Névot, P. Croce, Caractérisation des surfaces par réflexion rasante de rayons X. Application à l'étude du polissage de quelques verres silicates, Rev. Phys. Appl. (Paris), 15 (1980) 761-779.

[38] A.R. Rennie, http://www.reflectometry.net/reflect.htm#Analysis.

[39] S. Basu, S. Singh, A new polarized neutron reflectometer at Dhruva for specular and off-specular neutron reflectivity studies, Journal of Neutron Research, 14 (2006) 109-120.

[40] S. Singh, D. Bhattacharya, S. Basu, Specular and off-specular polarized neutron reflectivity at Dhruva, Neutron News, 25 (2014) 31-33.

[41] Z. Fisher, A. Jackson, A. Kovalevsky, E. Oksanen, H. Wacklin, Chapter 1 - Biological Structures, in: F. Fernandez-Alonso, D.L. Price (Eds.) Experimental Methods in the Physical Sciences, Academic Press2017, pp. 1-75.

[42] S. Mitragotri, D.G. Anderson, X. Chen, E.K. Chow, D. Ho, A.V. Kabanov, J.M. Karp, K. Kataoka, C.A. Mirkin, S.H. Petrosko, J. Shi, M.M. Stevens, S. Sun, S. Teoh, S.S. Venkatraman, Y. Xia, S. Wang, Z. Gu, C. Xu, Accelerating the Translation of Nanomaterials in Biomedicine, ACS Nano, 9 (2015) 6644-6654.

[43] K. Simons, M.J. Gerl, Revitalizing membrane rafts: new tools and insights, Nature Reviews Molecular Cell Biology, 11 (2010) 688-699.

[44] P.L. Yèagle, Lipid regulation of cell membrane structure and function, The FASEB Journal, 3 (1989) 1833-1842.

[45] G. Fragneto, Neutrons and model membranes, The European Physical Journal Special Topics, 213 (2012) 327-342.

[46] G. Fragneto, R. Delhom, L. Joly, E. Scoppola, Neutrons and model membranes: Moving towards complexity, Current Opinion in Colloid & Interface Science, 38 (2018) 108-121.

[47] G. Fragneto, T. Charitat, F. Graner, K. Mecke, L. Perino-Gallice, E. Bellet-Amalric, A fluid floating bilayer, Europhys. Lett., 53 (2001) 100-106.

[48] A.R. Mazzer, L.A. Clifton, T. Perevozchikova, P.D. Butler, C.J. Roberts, D.G. Bracewell, Neutron reflectivity measurement of protein A–antibody complex at the solid-liquid interface, Journal of Chromatography A, 1499 (2017) 118-131.

[49] A. Martel, L. Antony, Y. Gerelli, L. Porcar, A. Fluitt, K. Hoffmann, I. Kiesel, M. Vivaudou, G. Fragneto, J.J. de Pablo, Membrane Permeation versus Amyloidogenicity: A Multitechnique Study of Islet Amyloid Polypeptide Interaction with Model Membranes, Journal of the American Chemical Society, 139 (2017) 137-148.

[50] N. Banu, S. Singh, S. Basu, A. Roy, H.C.P. Movva, V. Lauter, B. Satpati, B.N. Dev, High density nonmagnetic cobalt in thin films, Nanotechnology, 29 (2018) 195703.

[51] M.A. Basha, C.L. Prajapat, M. Gupta, H. Bhatt, Y. Kumar, S.K. Ghosh, V. Karki, S. Basu, S. Singh, Interface induced magnetic properties of Gd/Co heterostructures, Physical Chemistry Chemical Physics, 20 (2018) 21580-21589.

[52] S. Singh, M.A. Basha, H. Bhatt, Y. Kumar, M. Gupta, Interface morphology driven exchange interaction and magnetization reversal in a Gd/Co multilayer, Physical Chemistry Chemical Physics, 24 (2022) 6580-6589.

[53] J. Hoppler, J. Stahn, C. Niedermayer, V.K. Malik, H. Bouyanfif, A.J. Drew, M. Rössle, A. Buzdin, G. Cristiani, H.U. Habermeier, B. Keimer, C. Bernhard, Giant superconductivity-induced modulation of the ferromagnetic magnetization in a cuprate–manganite superlattice, Nature Materials, 8 (2009) 315-319.

[54] D.K. Satapathy, M.A. Uribe-Laverde, I. Marozau, V.K. Malik, S. Das, T. Wagner, C. Marcelot, J. Stahn, S. Brück, A. Rühm, S. Macke, T. Tietze, E. Goering, A. Frañó, J.H. Kim, M. Wu, E. Benckiser, B. Keimer, A. Devishvili, B.P. Toperverg, M. Merz, P. Nagel, S. Schuppler, C. Bernhard, Magnetic Proximity Effect in $YBa_2Cu_3O_7/La_{2/3}Ca_{1/3}MnO_3$ and $YBa_2Cu_3O_7/LaMnO_{3+\delta}$ Superlattices, Physical Review Letters, 108 (2012) 197201.

[55] P. Yu, J.S. Lee, S. Okamoto, M.D. Rossell, M. Huijben, C.H. Yang, Q. He, J.X. Zhang, S.Y. Yang, M.J. Lee, Q.M. Ramasse, R. Erni, Y.H. Chu, D.A. Arena, C.C. Kao, L.W.

Martin, R. Ramesh, Interface Ferromagnetism and Orbital Reconstruction in BiFeO$_3$-La$_{0.7}$Sr$_{0.3}$MnO$_3$ Heterostructures, Physical Review Letters, 105 (2010) 027201.

[56] S. Singh, J. Xiong, A.P. Chen, M.R. Fitzsimmons, Q.X. Jia, Field-dependent magnetization of BiFeO$_3$ in an ultrathin La$_{0.7}$Sr$_{0.3}$MnO$_3$/BiFeO$_3$ superlattice, Physical Review B, 92 (2015) 224405.

[57] P. Jain, Q. Wang, M. Roldan, A. Glavic, V. Lauter, C. Urban, Z. Bi, T. Ahmed, J. Zhu, M. Varela, Q.X. Jia, M.R. Fitzsimmons, Synthetic magnetoelectric coupling in a nanocomposite multiferroic, Scientific Reports, 5 (2015) 9089.

[58] C.L. Prajapat, H. Bhatt, Y. Kumar, T.V.C. Rao, P.K. Mishra, G. Ravikumar, C.J. Kinane, B. Satpati, A. Caruana, S. Langridge, S. Basu, S. Singh, Interface-Induced Magnetization and Exchange Bias in LSMO/BFO Multiferroic Heterostructures, ACS Applied Electronic Materials, 2 (2020) 2629-2637.

[59] J. Wang, J.B. Neaton, H. Zheng, V. Nagarajan, S.B. Ogale, B. Liu, D. Viehland, V. Vaithyanathan, D.G. Schlom, U.V. Waghmare, N.A. Spaldin, K.M. Rabe, M. Wuttig, R. Ramesh, Epitaxial BiFeO3 Multiferroic Thin Film Heterostructures, Science, 299 (2003) 1719-1722.

[60] C. Ederer, N.A. Spaldin, Weak ferromagnetism and magnetoelectric coupling in bismuth ferrite, Physical Review B, 71 (2005) 060401.

[61] C. He, A.J. Grutter, M. Gu, N.D. Browning, Y. Takamura, B.J. Kirby, J.A. Borchers, J.W. Kim, M.R. Fitzsimmons, X. Zhai, V.V. Mehta, F.J. Wong, Y. Suzuki, Interfacial Ferromagnetism and Exchange Bias in CaRuO$_3$/CaMnO$_3$ Superlattices, Physical Review Letters, 109 (2012) 197202.

[62] T.S. Santos, B.J. Kirby, S. Kumar, S.J. May, J.A. Borchers, B.B. Maranville, J. Zarestky, S.G.E. te Velthuis, J. van den Brink, A. Bhattacharya, Delta Doping of Ferromagnetism in Antiferromagnetic Manganite Superlattices, Physical Review Letters, 107 (2011) 167202.

[63] P.A. Salvador, A.M. Haghiri-Gosnet, B. Mercey, M. Hervieu, B. Raveau, Growth and magnetoresistive properties of (LaMnO$_3$)$_m$(SrMnO$_3$)$_n$ superlattices, Applied Physics Letters, 75 (1999) 2638-2640.

[64] A.J. Grutter, H. Yang, B.J. Kirby, M.R. Fitzsimmons, J.A. Aguiar, N.D. Browning, C.A. Jenkins, E. Arenholz, V.V. Mehta, U.S. Alaan, Y. Suzuki, Interfacial Ferromagnetism in LaNiO$_3$/CaMnO$_3$ Superlattices, Physical Review Letters, 111 (2013) 087202.

[65] J. Hoffman, I.C. Tung, B.B. Nelson-Cheeseman, M. Liu, J.W. Freeland, A. Bhattacharya, Charge transfer and interfacial magnetism in (LaNiO$_3$)$_n$/(LaMnO$_3$)$_2$ superlattices, Physical Review B, 88 (2013) 144411.

[66] F. Katmis, V. Lauter, F.S. Nogueira, B.A. Assaf, M.E. Jamer, P. Wei, B. Satpati, J.W. Freeland, I. Eremin, D. Heiman, P. Jarillo-Herrero, J.S. Moodera, A high-temperature ferromagnetic topological insulating phase by proximity coupling, Nature, 533 (2016) 513-516.